\documentclass[twocolumn,showpacs,aps,superscriptaddress,%
  eqsecnum,prd,notitlepage,showkeys,10pt]{revtex4-1}

\usepackage{amssymb}
\usepackage{amsmath}
\usepackage{graphicx}
\usepackage{dcolumn}
\usepackage{hyperref}
\usepackage{upgreek}

\begin{document}

\title{First Search for the Associated Production \endgraf of a Higgs Boson with a Single Top Quark}

\author{Benedikt Maier}
\affiliation{Karlsruhe Institute of Technology (KIT) -- D-76131 Karlsruhe, Germany \\ \vspace{-0.3cm}${}$\\on behalf of the CMS collaboration.\\\vspace{0.2cm} ${}$\\Preprint of the proceedings for the contribution to the LHCP2015 conference, St.\,Petersburg, Russia.}

\begin{abstract}
The production of the Higgs boson in association with a single top quark is sensitive to the relative sign of the coupling parameters describing its interaction with fermions and gauge bosons. The tHq production mode therefore provides an good handle on the Yukawa coupling $Y_\mathrm{t}$. The first searches for single-top + Higgs in the $\mathrm{H}\to \mathrm{b}\bar{\mathrm{b}}$, $\upgamma\upgamma$, $\uptau^+\uptau^-$ and $\mathrm{W}^+\mathrm{W}^-$ decay channels are presented, using the full 8\,TeV dataset recorded with the CMS detector. Special emphasis is put on the analyses' peculiarities and their dominating systematic uncertainties, and a combination of all individual channels is performed. The analyses are optimized for a scenario of $Y_\mathrm{t}=-1$, which is enhanced by a factor of $\sim13$ with respect to the Standard Model production rate. The observed combined upper exclusion limit is $2.8\times\sigma_{Y_\mathrm{t}=-1}$ ($2.0$ expected).
\end{abstract}

\maketitle

\section{Introduction}

In 2012 \cite{ATLAS,CMS}, the ATLAS and CMS collaborations announced the discovery of a new boson that is consistent with the Higgs boson as postulated in the 1960's~\cite{englert,higgs}. Since then it is the goal to measure its characteristics as precisely as possible in order to pin down possible deviations from the Standard Model (SM) predictions. The Yukawa coupling mechanism to fermions is an important feature and subject to such tests. In the theory coupling strengths are proportional to the fermion masses. In particular, since the top quark is the heaviest elementary particle known to exist, the coupling $Y_\mathrm{t}$ is a significant parameter for verification of the electroweak sector of the SM. According to $Y_\mathrm{f}=\sqrt{2}\frac{m_\mathrm{f}}{v}$, where the vacuum expectation value of the Higgs field is $v\sim246$\,GeV, this gives an absolute value of $Y_\mathrm{t}\simeq 1$. This is in accordance with recent measurements \cite{ATLAS2,CMS2}. Most channels however are insensitive to the sign of $Y_\mathrm{t}$ or, more precisely, to its relative sign with respect to the parameter describing the coupling of the Higgs boson to gauge bosons.

\begin{figure}
  \includegraphics[width=.185\textwidth]{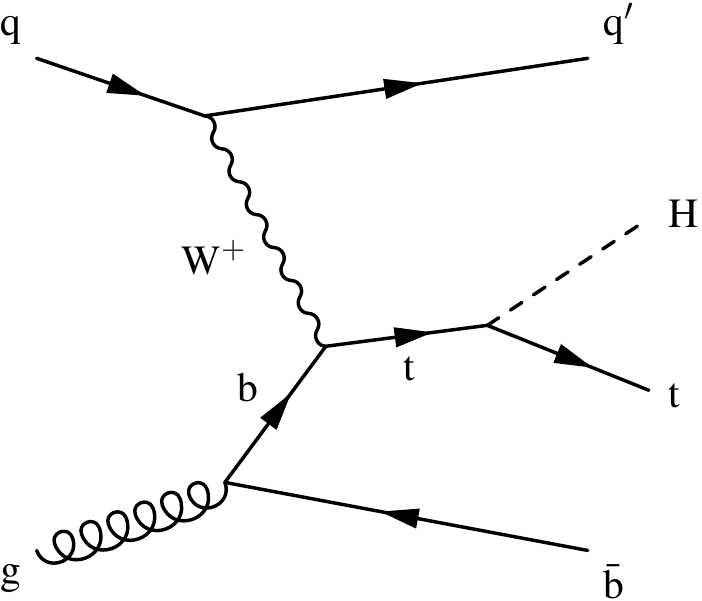}\hspace{0.4cm}
  \includegraphics[width=.185\textwidth]{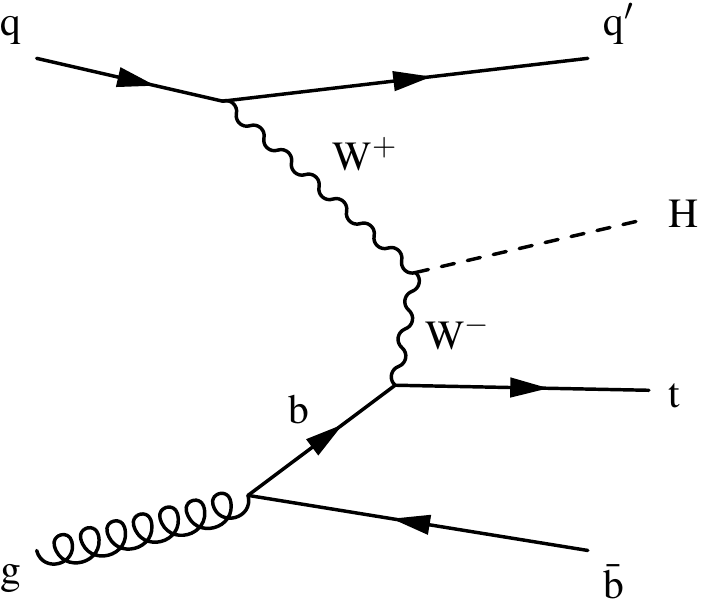}\\
 \caption{Leading order tHq Feynman diagrams.}
 \label{fig:feyns}
\end{figure}

Figure~\ref{fig:feyns} shows two leading-order Feynman diagrams for the associated production of a Higgs boson with a single top quark in the $t$-channel.\footnote{Diagrams where the Higgs boson is attached to b quark lines can be neglected because their contribution is suppressed by $(m_\mathrm{b}/m_\mathrm{t})^2$.} 
For the SM, there exists a destructive interference between the two diagrams, which results in a tiny production cross section of $\sigma_{\mathrm{tHq}}=18.3$\,fb \cite{farina}. The scenario of $C_\mathrm{t}=-1$ has a cross section enhanced by a factor $\sim 13$. This brings it into reach for searches with the integrated luminosity collected at $\sqrt{s}=8$\,TeV. First and most recent phenomenological studies on tHq production can be found in \cite{maltoni,demartin}.

\section{Channel topology}

For a single top $t$-channel-like process, the most characteristic feature is the upper outgoing quark line in Figure~\ref{fig:feyns}, which represents a light quark that has recoiled against the exchanged virtual W boson. It produces a typically very forward light jet with a substantial $p_\mathrm{T}$.  The other resonance besides the Higgs boson, the top quark, is required to decay leptonically for all decay channels considered here. The presence of a prompt lepton can help in rejecting multi-jet background processes. Moreover, the sign of the lepton will be used for constructing same-sign final states together with leptons stemming from the Higgs boson for the relevant decay modes. The top decay also features a b~quark giving rise to a central b~jet. 
The initial gluon splitting creates a second, additional b~quark. The corresponding b jet however lies out of the tracker acceptance most of the time and thus cannot be tagged. All considered final states have to fight a large $\mathrm{t}\bar{\mathrm{t}}+X$ background ($X=W,Z,H,\mathrm{or\,jets}$).

\newcommand{\pt}{\ensuremath{p_\mathrm{T}}}
\newcommand{\ttbar}{\ensuremath{\mathrm{t}\bar{\mathrm{t}}}}
\newcommand{\ttbarH}{\ensuremath{\mathrm{t}\bar{\mathrm{t}}\mathrm{H}}}
\newcommand{\abseta}{\ensuremath{\mid\hspace{-0.085cm}\eta\hspace{-0.085cm}\mid\,\,}}

\section{Higgs Boson final states}
\subsection{A pair of b~quarks}

Given the small production rates, the decay $\mathrm{H}\to \mathrm{b}\bar{\mathrm{b}}$ with a branching fraction of $58\%$ is a promising channel, as it retains most of the anyways sparse signal events. The electron (muon) from the leptonic top decay is required to have a transverse momentum larger than 30\,GeV (26\,GeV) and to lie in a central detector region with $\abseta<2.4$ ($2.1$). Additional leptons with a relaxed selection are vetoed in each event, leading to the rejection of Drell-Yan + jets processes. The analysis uses a jet \pt threshold of $\pt>20$\,GeV for central jets and 40\,GeV for forward jets. $E_\mathrm{T}^\mathrm{miss}$, which is identified with the escaping neutrino, is required to be $>45/35$\,GeV (e/$\upmu$). At least one untagged jet is required in the event selection. Two categories are introduced, differing in the number of b~tagged jets. The 3 tag category expects b~jets stemming from the decays of the two reconances. A 4 tag category is introduced to be sensitive to the fraction of events where the additional b~quark is produced centrally. The expected signal-over-background ($S/B$) ratios are $13/1900$ in the 3 tag region and $1.4/66$ in the 4 tag region.

\begin{figure}
\includegraphics[width=0.48\textwidth]{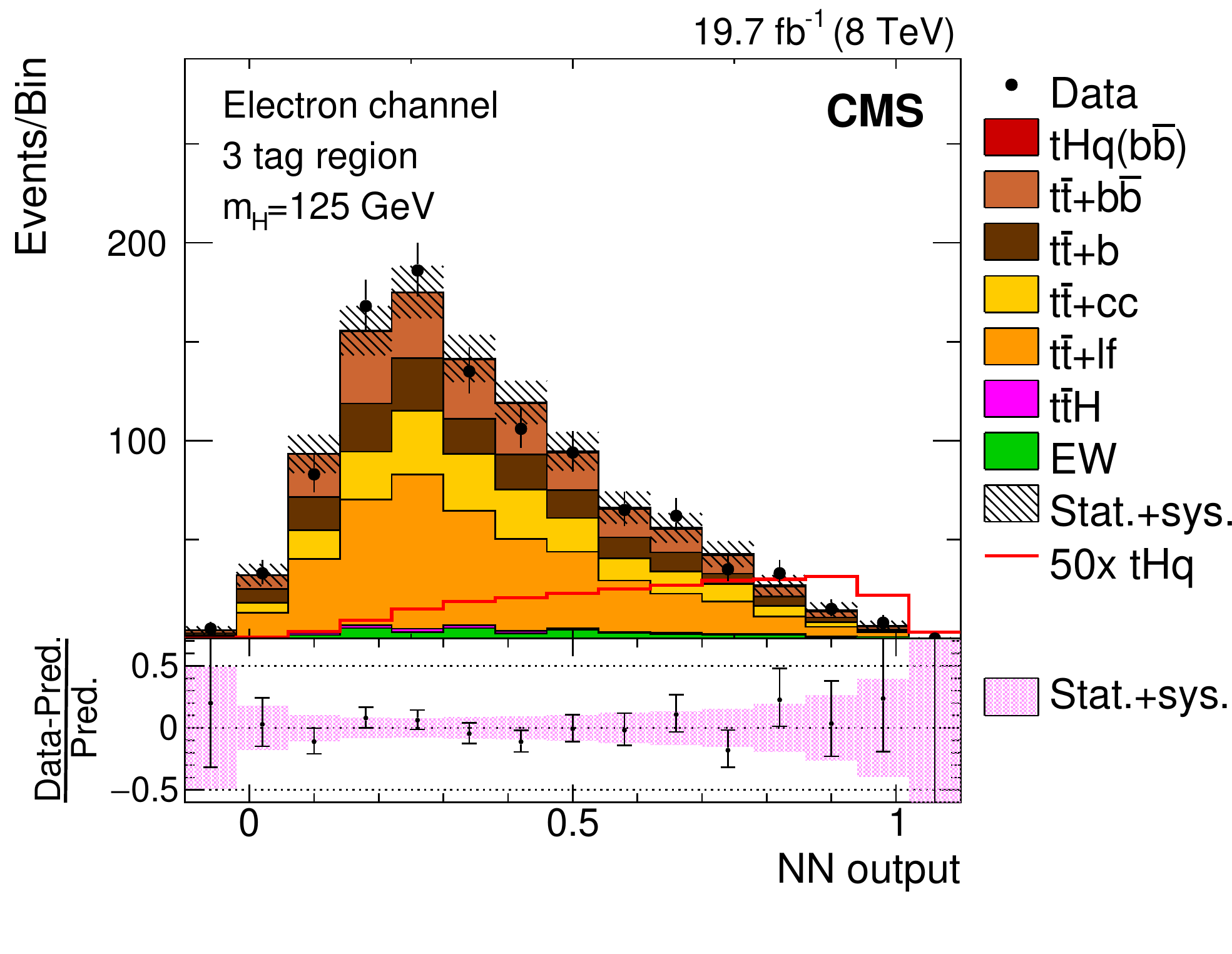}
\includegraphics[width=0.48\textwidth]{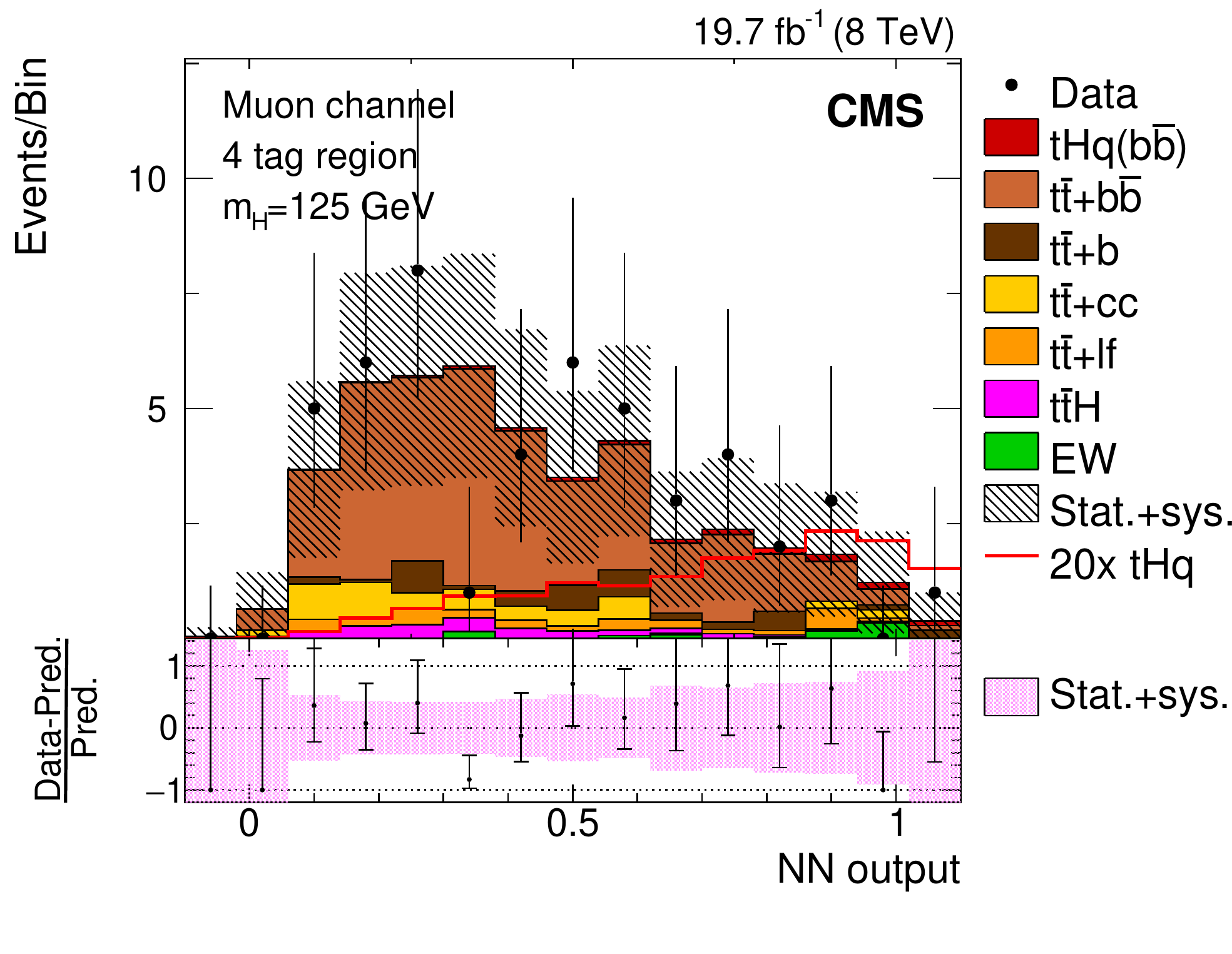}\\
 \caption{Post-fit NN output for the 3 tag region in the $\upmu$ channel (top) and the 4 tag region in the electron channel (bottom). The red hollow line gives the pre-fit expectation for tHq ($C_\mathrm{t}=-1$), scaled by a factor 50 (20).}
 \label{fig:bb}
\end{figure}

This means that even after a dedicated event selection as described above, there is a dominating background contribution mainly from \ttbar + jets production. A classification Neural Network (NN) is therefore employed to further separate the signal process from backgrounds, using as inputs observables that are genuine to tHq or $\mathrm{t}\bar{\mathrm{t}}$ events. Prior to this, a correspondence between the observed jets and the final state objects must be constructed in order to define the input variables to the classification NN in the most efficient way. Because of the large jet multiplicity, a correct jet assignment is a complex problem and is addressed by another Neural Network. It is trained with correct versus wrong jet assignments, where ``correct" refers to the event interpretation where each parton (the three b quarks from the resonances, and the light quark) can be matched uniquely to a reconstructed jets, and a ``wrong" interpretation is any other random jet assignment. When applying this reconstruction NN to unknown events, the event interpretation is picked that results in the largest response value of the discriminator. The same is done under the assumption the jets come from semi-leptonic \ttbar~production, matching the two b~quarks from the tops and the two light quarks from the hadronically decaying W boson. Based upon these interpretations, the final classification NN is fed with input variables such as the \pt~of the reconstructed Higgs boson, the mass of the reconstructed hadronically decaying W, and the lepton charge. The latter is an example for a variable that is independent from any type of reconstruction, but still provides a significant discrimination power between the symmetric case of \ttbar, and the $t$-channel-like tHq, which is more likely to be induced by quarks than by antiquarks, and consequently the charge of the lepton is $\sim$ twice more often positive than negative in proton-proton collisions.\footnote{In fact, the $t$-channel's high sensitivity to the quark/antiquark density in protons makes it an ideal place to constrain PDFs. This is being done for the case without the Higgs, where cross sections are much higher and almost pure signal samples can be obtained.}

Templates in the NN discriminator are then used to extract the signal and to set upper limits on $\sigma_\mathrm{tHq}$. Figure~\ref{fig:bb} shows the NN output distributions in two of the four analysis bins. The \ttbar + jets background has been split into categories varying in their additional heavy flavor content. The uncertainties on their rates and on higher order effects are the main sources of systematic uncertainty. An upper limit of $5.4\times\sigma_{C_\mathrm{t}=-1}$ at 95\% confidence level (C.L.) is found. The observed limit is slightly higher ($7.6$).

\subsection{Two photons}
\begin{figure}
\includegraphics[width=0.4\textwidth]{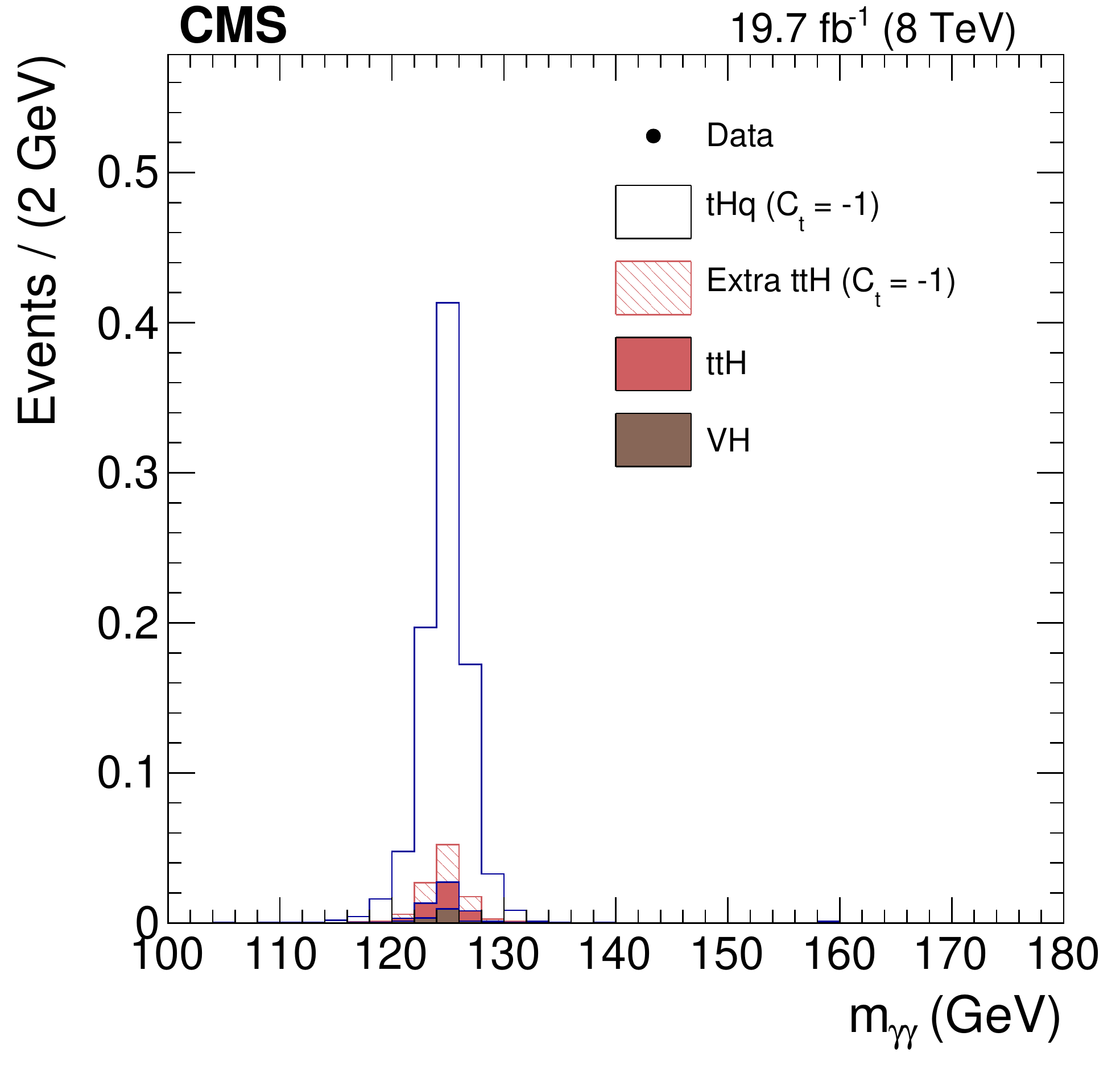}
\caption{Invariant mass of the reconstructed diphoton system. Zero evets are observed.}
 \label{fig:gg}
\end{figure}

The decay of the Higgs boson to two photons happens via a virtual loop of either top quarks or W bosons. Just like for the production there is a constructive interference for $C_\mathrm{t}=-1$, leading to a further enhancement on the expected rates by a factor 2.8. The event selection foresees two photons with $p_{\mathrm{T},\upgamma_1}>50 m_{\upgamma\upgamma}/120$\,GeV and $p_{\mathrm{T},\upgamma_2}>25$\,GeV, where $m_{\upgamma\upgamma}$ denotes the invariant mass of the reconstructed diphoton system. Further required are a b~tagged jet ($\pt>20$\,GeV), an untagged forward jet ($\pt>20$\,GeV and $\abseta>1$) and an isolated muon or electron with $\pt>10$\,GeV. The signal region is defined in the window $122<m_{\upgamma\upgamma}<128$\,GeV, i.e. $\pm 3$\,GeV around the expected Higgs mass. 

With several variables discriminating against \ttbarH, like the jet multiplicity in the event, the pseudorapidity of the light quark candidate jet or again the lepton charge, a simple likelihood classifier is constructed and cut on in order to reduce the resonant background. The resulting expected yields are $0.67$ for tHq, and $0.03+ 0.05$ for \ttbarH~and $0.01+ 0.01$ for VH. The latter numbers describe the aforementioned effect of enhanced rates in the Higgs boson decay. Predictions for all the Higgs related processes are taken from simulation. Other background contributions would stem from \ttbar~+ $\upgamma\upgamma$ or $\upgamma\upgamma$ + jets production. These have a non-resonant shape in the invariant diphoton mass $m_{\upgamma\upgamma}$ and are best determined with a data-driven technique from the sidebands $(100,122)$\,GeV and $(128,180)$\,GeV. In order to have enough data in these regions, b~tagging criteria are relaxed. A falling exponential is the assumed fuction\footnote{The uncertainty on the knowledge of the background shape is estimated using another control region with inverted isolation requirements on one of the two photons.}; its parameters are determined from the sidebands, and it is extrapolated into the signal region to estimate the contribution of the non-resonant backgrounds. 

In Figure~\ref{fig:gg} the resulting distribution for $m_{\upgamma\upgamma}$ is shown. Zero events are observed in both signal and sideband regions. In such a case the observed and expected limits coincide; the analysis is able to exclude $4.1\times\sigma_{C_\mathrm{t}=-1}$ at 95\% C.L.

\subsection{Multi-leptons}

In the trilepton channel contributions are expected from events where the Higgs boson decayed into a pair of W bosons or taus which then have an entirely leptonic decay chain. This leads to the allowed lepton combinations (eee), ($\upmu\upmu\upmu$), (ee$\upmu$) and (e$\upmu\upmu$) with $\pt>20/10/10$\,GeV. A cut on $E_\mathrm{T}^\mathrm{miss}>30$\,GeV accounts for the presence of three neutrinos. The reconstructed dilepton mass closest to $m_\mathrm{Z}$ must lie outside ($m_\mathrm{Z}\pm15$\,GeV) to suppress the Drell-Yan background. Exactly one central jet must be tagged as b~jet, and at least one forward jet is required with $\abseta>1.5$. The dilepton channel asks for exactly two leptons with same electric charge, allowing the combinations (e$\upmu$), ($\upmu\upmu$) with $p_{\mathrm{T},\ell}>25$\,GeV and $m_{\ell\ell}>20$\,GeV. \begin{figure}[h!]
\includegraphics[width=0.28\textwidth]{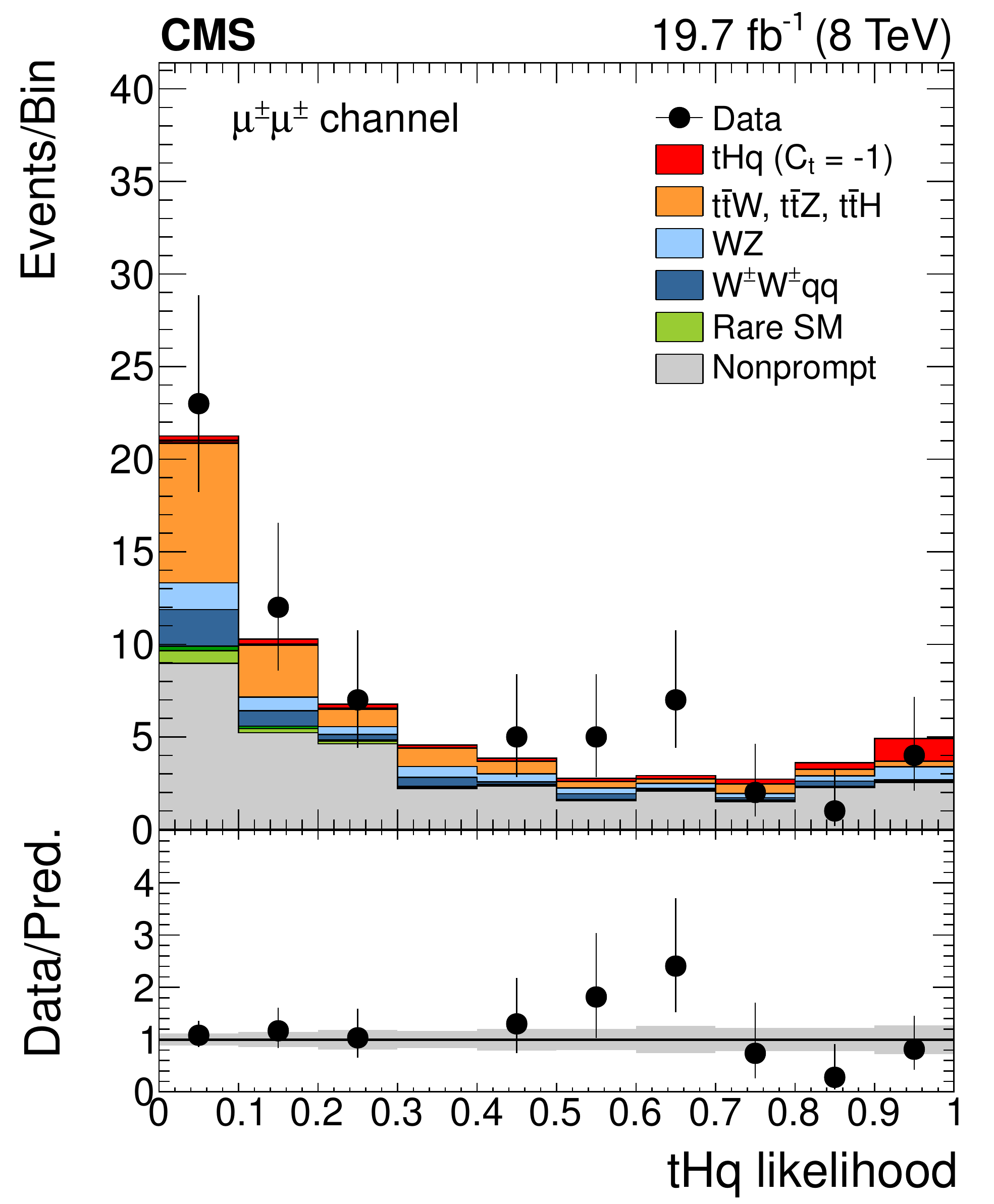}
\includegraphics[width=0.28\textwidth]{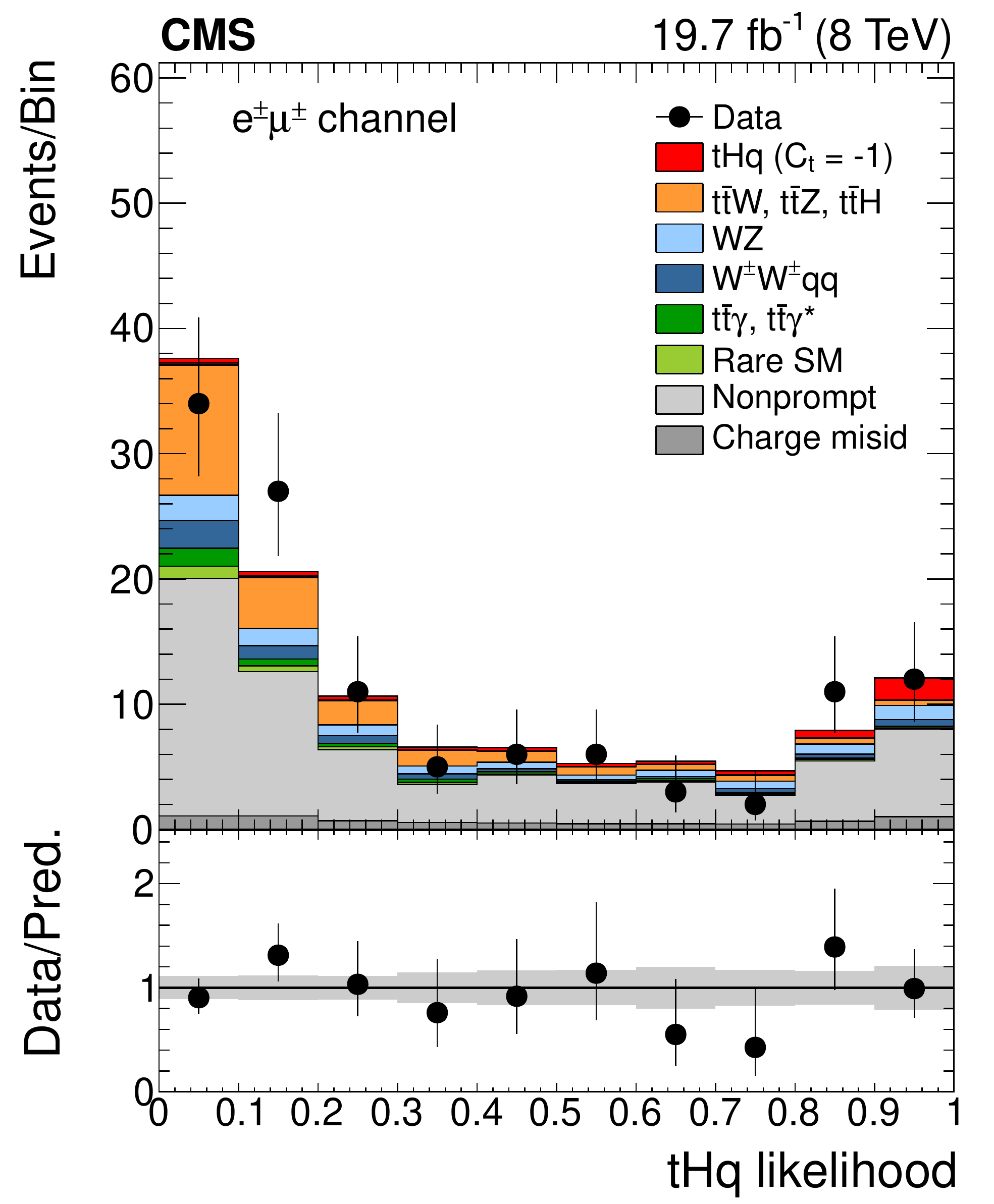}
\includegraphics[width=0.28\textwidth]{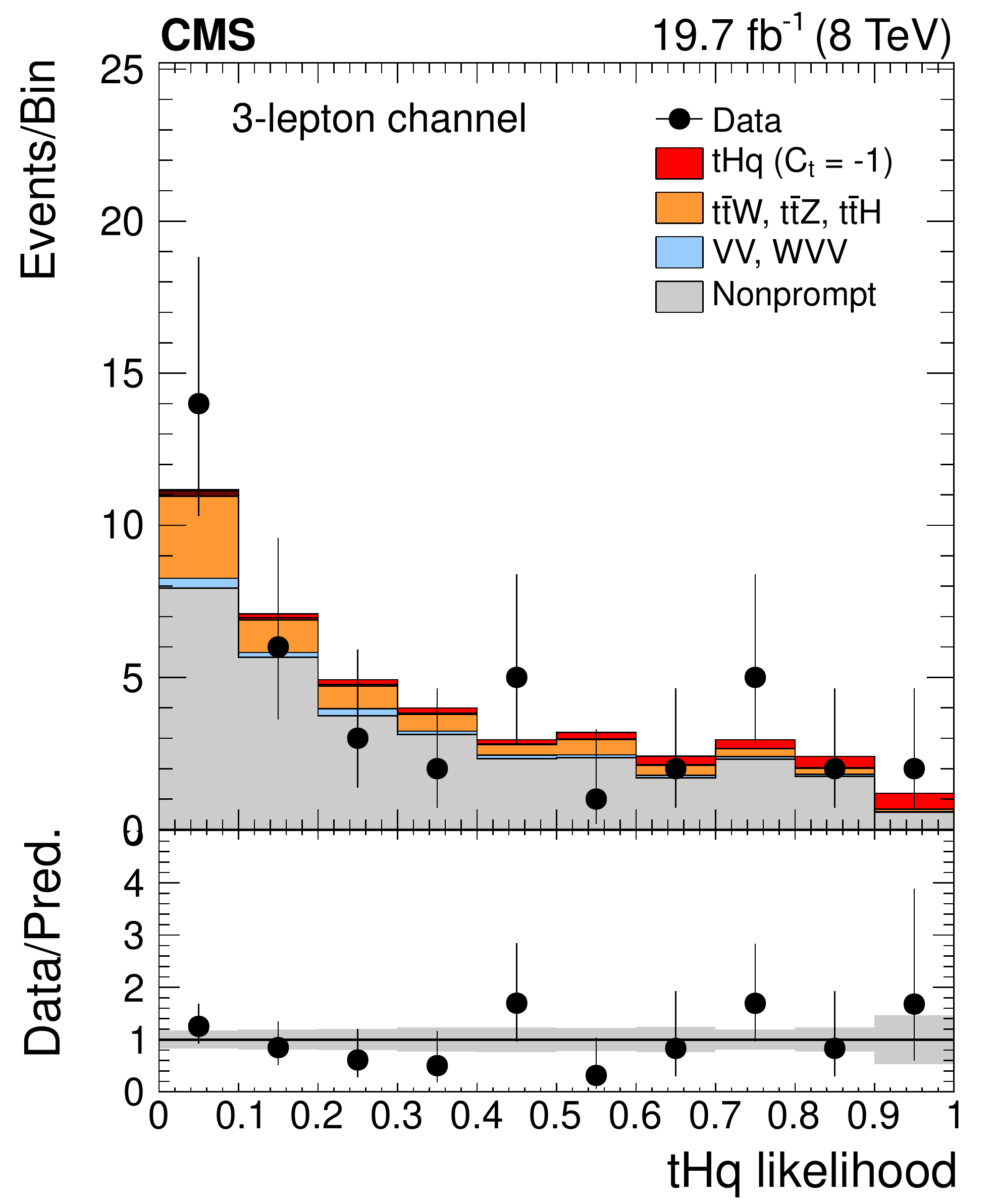}
\caption{Post-fit likelihood disciminator for the $\upmu\upmu$ (top), the e$\upmu$ (center) and the trilepton channel (bottom). The gray bands represent the combined statistical and systematic uncertainties as determined in the  maximum-likelihood fit to data.}
 \label{fig:ww}
\end{figure}

There has to be at least one central tagged b~jet and one or more forward jets ($\abseta>1.0$). Since for this channel one of the W bosons is assumed to decay hadronically, one additional central jet is required. Hadronically decaying $\uptau$'s are vetoed explicitly. The jet \pt~threshold for both channels is 25\,GeV.

The most significant background comes from \ttbar~events, where leptons can be produced in the decay of B hadrons, or when light jets are misidentified as leptons. A ``tight-to-loose" method employs the \pt- and $\eta$-dependent probabilities that a non-prompt lepton which passes looser isolation and impact parameter criteria also fulfills the tight lepton ID criteria used in the analysis. It estimates the probabilities in data using a control sample enriched in background leptons. With the determined fake rates the event yields in sideband regions differing only in the lepton isolation can be weighted into the signal region to obtain an estimate for the non-prompt backgrounds.  Contamination due to misidentified lepton charge is estimated from $\mathrm{Z}\to\ell\ell$ events. The misidentification rate for electrons amounts to $<0.08\%$ in the barrell and $\sim 0.28\%$ in the endcap. For muons it is negligible. A likelihood discriminator is built from information on lepton charge and kinematics, forward jet activity and b~jet multiplicity. It discriminates between 3.3 (2.6) signal events and 106 (53) background events for the e$\upmu$ ($\upmu\upmu$) channel. The $S/B$ ratio for the trilepton channel is $1.5/42$. The post-fit classifier output in all channels can be seen in Figure~\ref{fig:ww}. Upper exclusion limits at 95\% C.L. on tHq production are derived from these distributions and is found to be $5.0\times\sigma_{C_\mathrm{t}=-1}$ (expected) and $6.7\times\sigma_{C_\mathrm{t}=-1}$ (observed), respectively.

\subsection{\boldmath$\uptau_\mathrm{lep}$  \boldmath$\uptau_\mathrm{had}$}

While there is a substantial leakage of events with two leptonically decaying taus into the previously described selection of the multi-lepton search, this complementary analysis is looking for final states where a hadronically decaying $\uptau$ could be reconstructed and two other same-sign leptons (e$\upmu$,$\upmu\upmu$) are identified, one of which is expected to stem from the top quark decay. The same-sign requirement strongly suppresses backgrounds with a prompt dilepton pair of opposite charge, like in $\mathrm{Z}/\upgamma^*\to\upmu\upmu$, that has been produced in association with a faked hadronically decaying $\uptau$. The two leading leptons (electrons or muons) must fulfill $\pt>20/10$\,GeV. For electrons (muons), the $\eta$ requirement is $\abseta<2.5$ (2.4). A boosted decision tree trained with variables affecting lepton isolation is used to further reject events with secondary leptons such as from B hadron decays. The third lepton -- the hadronically decaying tau, $\uptau_\mathrm{had}$ -- must have a transverse momentum larger than $20$\,GeV and must be reconstructed in a central detector region with $\abseta<2.3$. It must have opposite charge compared to the other two leptons. All three of them need be separated by $\Delta R_{\ell\ell}>0.5$. At least one b~tagged jet is required with $\pt>25$\,GeV. This does not only reflect the expected signal topology with the top quark decay, but also significantly reduces the $\mathrm{Z}\to\uptau\uptau$ + jets background, which lacks a genuine b jet.

\begin{figure}
  \includegraphics[width=.38\textwidth]{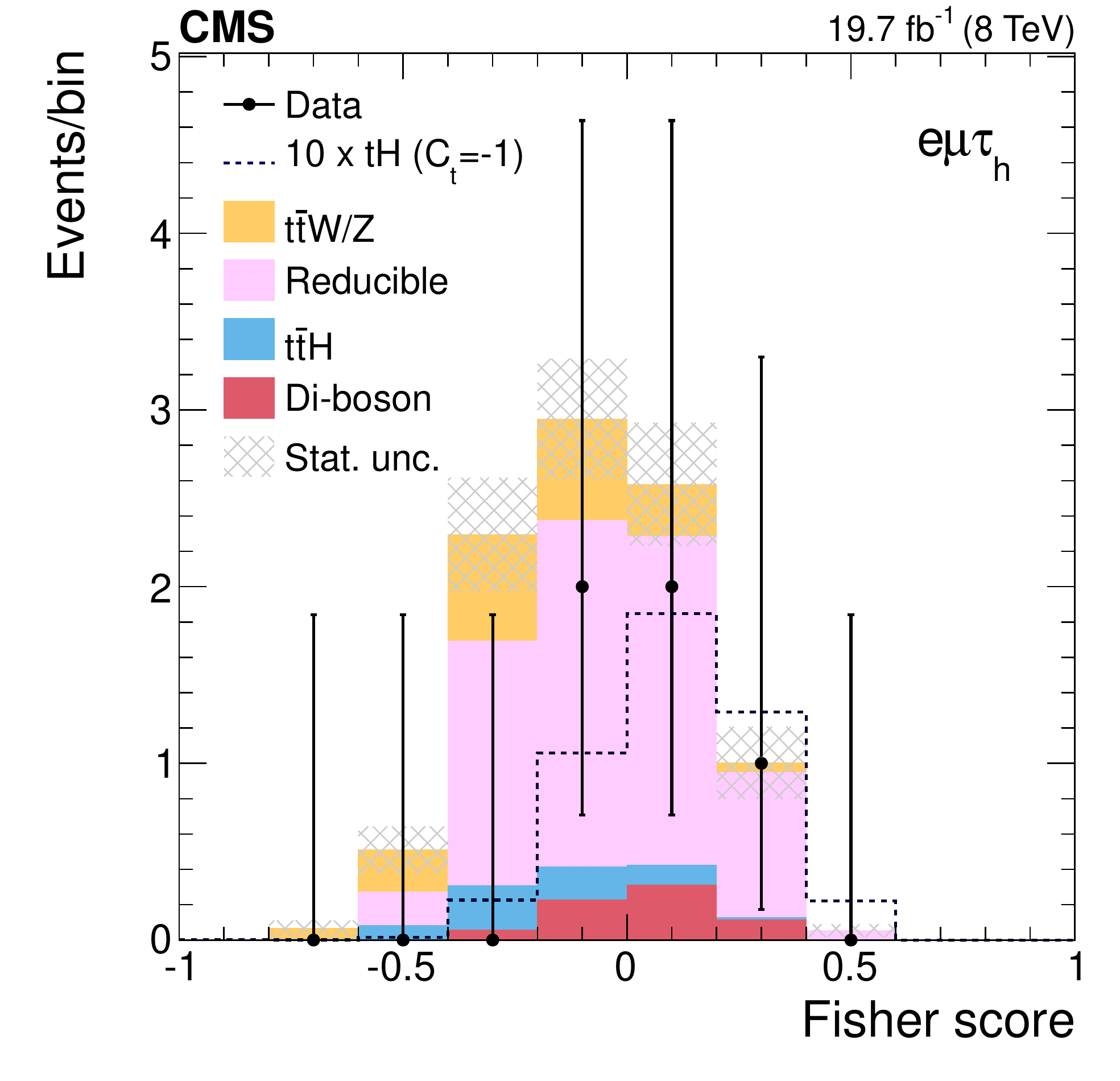} 
\hspace{15mm}
  \includegraphics[width=.38\textwidth]{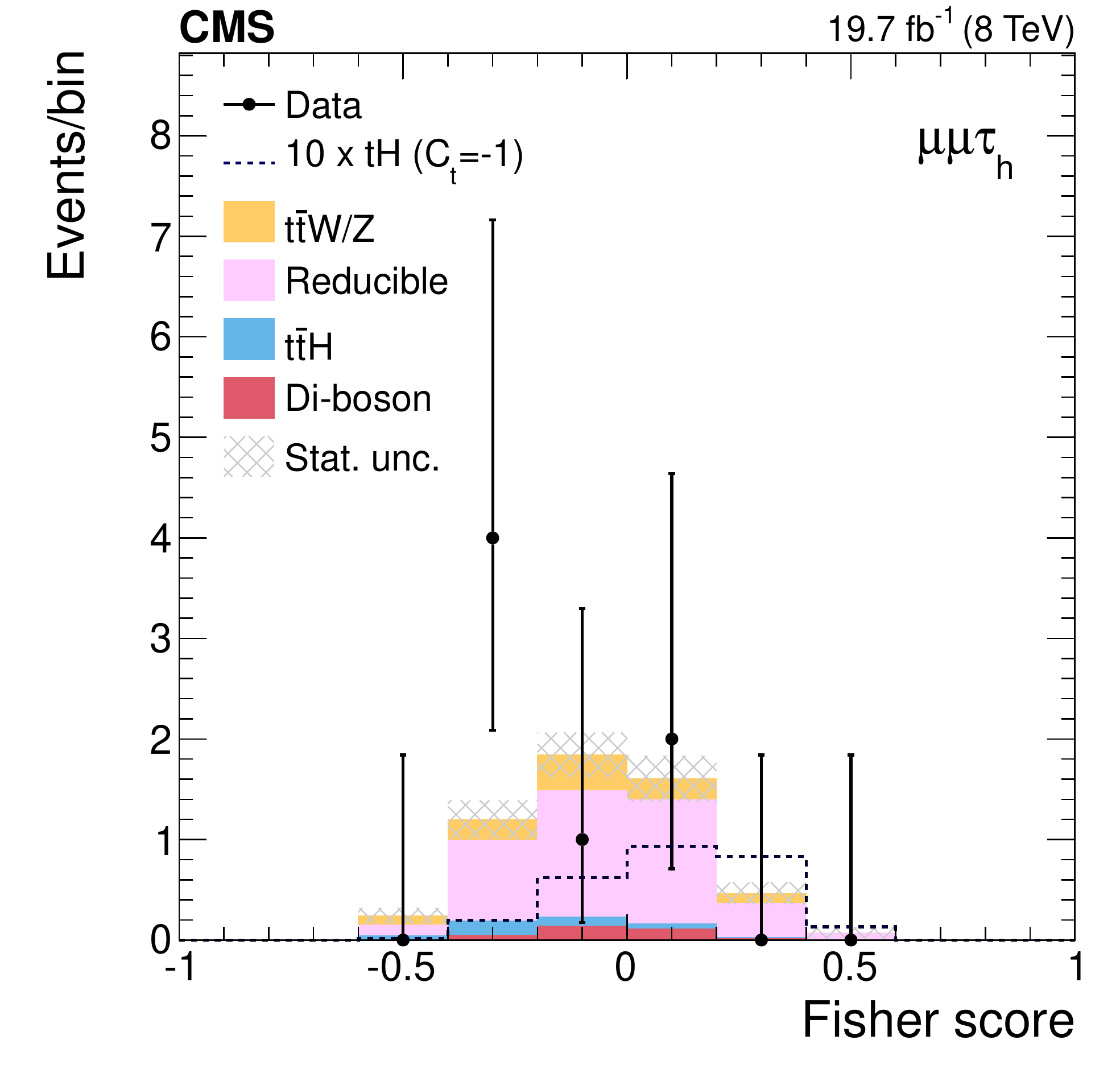}\\
 \caption{Post fit Fisher discriminant distributions for the e$\upmu$ (top) and $\upmu\upmu$ (bottom) channel. The dotted line gives the expected contribution from tHq; for making it visible it is scaled up by a factor 10.}
 \label{fig:tautau}
\end{figure}

Akin to the situation in the multi-lepton analysis, background contributions due to misidentified non-prompt leptons are estimated using a data-driven technique via fake rates in control samples and applying them to a signal sideband region. Irreducible backgrounds such as diboson production or \ttbar~+W/Z are modelled using Monte-Carlo simulations. One expects 0.48 (0.30) signal events and 9.5 (5.4) background events in the e$\upmu\uptau_\mathrm{had}$ ($\upmu\upmu\uptau_\mathrm{had}$) channel. A Fisher discriminant shall separate between tHq and the backgrounds. It is formed from variables describing e.g. the forward jet kinematics and the b~jet multiplicity. The training is performed in a control region with inverted isolation criteria on the reconstructed $\uptau_\mathrm{had}$ because of statistics.

The expected upper limit is derived from the two Fisher discriminators in Figure~\ref{fig:tautau} and is (at 95\% C.L.) $11\times\sigma_{C_\mathrm{t}=-1}$, while the data allows to exclude scenarios with cross sections larger than $9\times\sigma_{C_\mathrm{t}=-1}$.

\section{Combination}

\begin{figure}
\includegraphics[width=0.4038\textwidth]{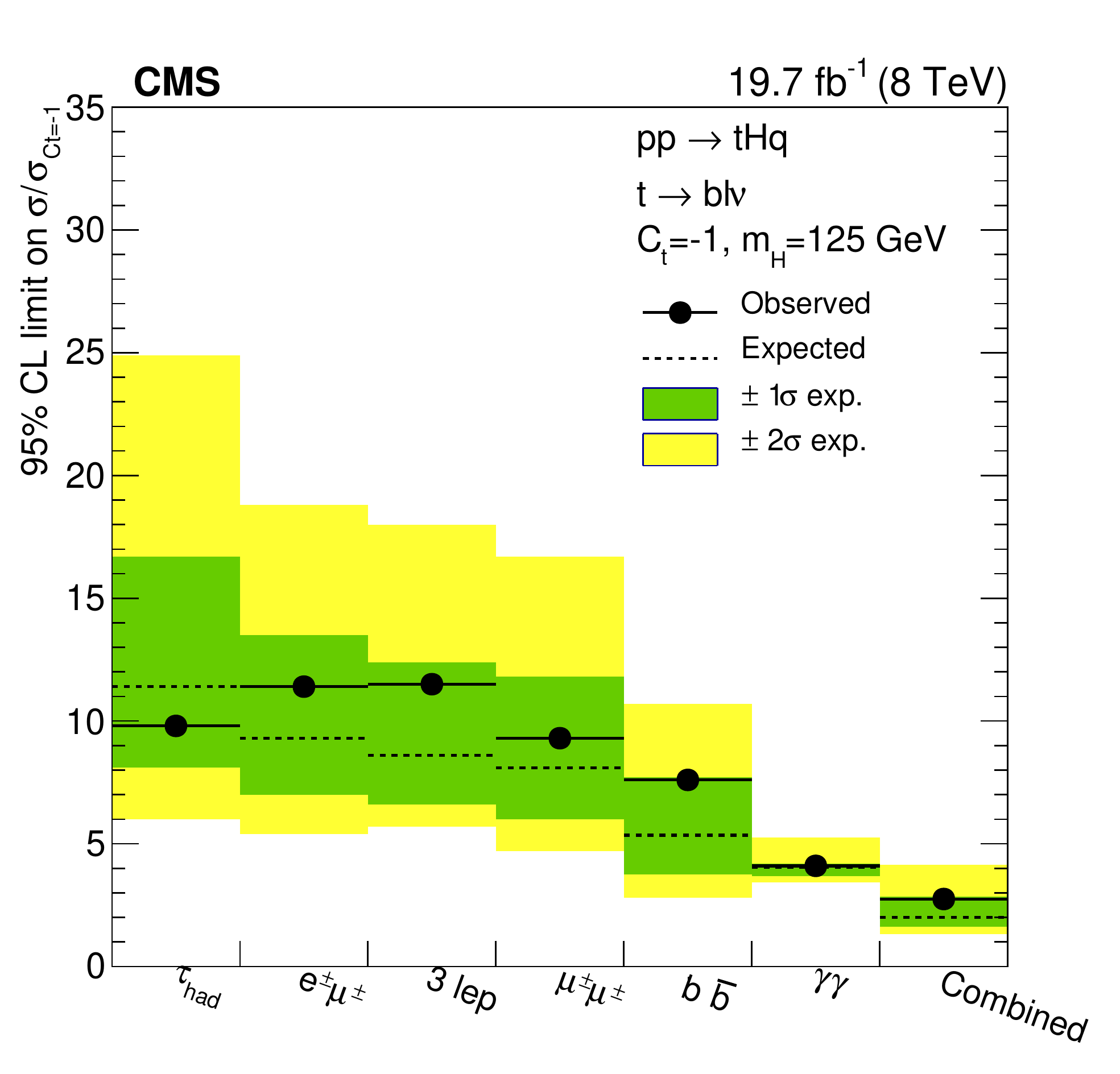}
\hspace{10mm}
\includegraphics[width=0.40\textwidth]{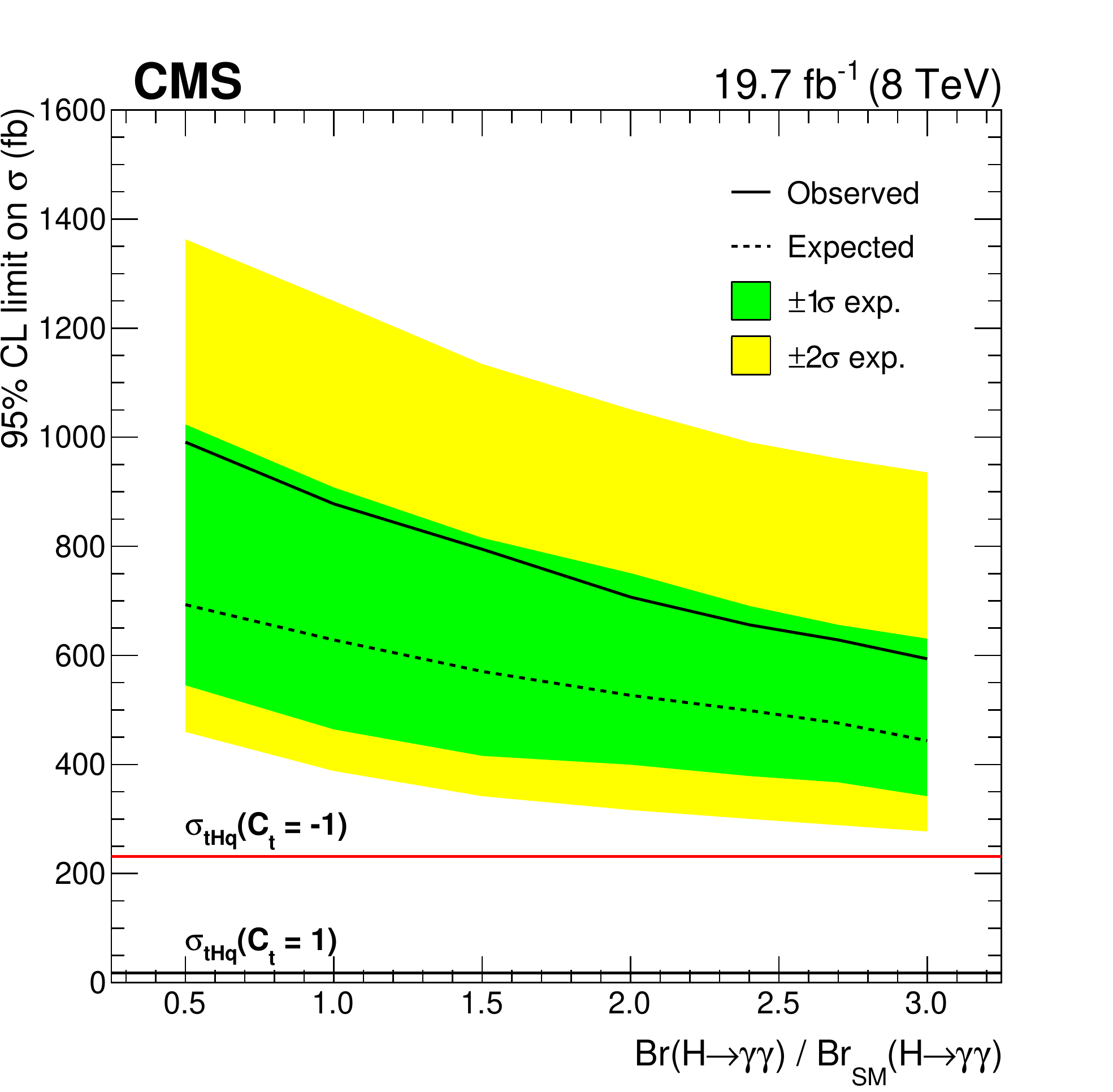}
\caption{Top: observed and expected exclusion limits for $C_\mathrm{t}=-1$ for the single analyses as well as for their combination. Here the additional expected contributions from $C_\mathrm{t}=-1$ in the decay $\mathrm{H}\to\upgamma\upgamma$ are explicitly taken into account as signal. Bottom: the search sensitivity is quoted as a function of $\mathcal{B}(\mathrm{H}\to\upgamma\upgamma)$.}
 \label{fig:comb}
\end{figure}

All of the distributions that have been shown so far can be used to derive common observed and expected upper limits on the $C_\mathrm{t}=-1$ scenario. The predictions of all channels are simultaneously fit to data; the underlying statistical model involves all the nuisance parameters of the single analyses. Two results are presented: the first approach fully takes the enhancement effects also in the decay to two photons into account. The expected search sensitivity turns out to be  $2.0\times\sigma_{C_\mathrm{t}=-1}$ at 95\% C.L., the observed limit is $2.8\times\sigma_{C_\mathrm{t}=-1}$. Another approach provides the limits as a function of the branching fraction $\mathcal{B}(\mathrm{H}\to\upgamma\upgamma)$, which depends on $C_\mathrm{t}$. Figure~\ref{fig:comb} summarizes the limits of the single analyses and shows the combined results.

\end{document}